\documentclass[reprint,showpacs,preprintnumbers,amsmath,amssymb]{revtex4-1}
\pdfoutput=1
\usepackage[pdftex]{graphicx}
\usepackage{color}
\usepackage{amsmath}
\usepackage{braket}
\usepackage[T1]{fontenc}
\usepackage{url}
\usepackage[english]{babel}
\usepackage{graphicx}
\usepackage{dcolumn}
\usepackage{bm}
\usepackage{subfig}
\usepackage{float}
\usepackage{url}

\newcommand{\Tr}{\mbox{\rm Tr}}
\newcommand{\ReC}{\mbox{\rm Re}}

\newcommand{\be}{\begin{equation}}
\newcommand{\ee}{\end{equation}}
\newcommand{\bea}{\begin{eqnarray}}
\newcommand{\eea}{\end{eqnarray}}
\newcommand{\non}{\nonumber}
\newcommand{\bit}{\begin{itemize}}
\newcommand{\eit}{\end{itemize}}
\newcommand{\mbf}{\mathbf}

\usepackage{verbatim}

\graphicspath{{figs/}}

\begin{document}

\title{Inside the SU(3) quark-antiquark QCD flux tube:  screening versus quantum widening}

\author{
N. Cardoso}
\email{nunocardoso@cftp.ist.utl.pt}
\author{
M. Cardoso}
\email{mjdcc@cftp.ist.utl.pt}
\author{
P. Bicudo}
\email{bicudo@ist.utl.pt}
\affiliation{CFTP, Departamento de F\'{\i}sica, Instituto Superior T\'{e}cnico,
Av. Rovisco Pais, 1049-001 Lisboa, Portugal}

\begin{abstract}
In lattice QCD,  colour confinement manifests in flux tubes.
We compute in detail the quark-antiquark flux tube for pure  gauge SU(3) dimension $D=3+1$ for quark-antiquark distances $R$ ranging from 0.4 fm to 1.4 fm.  
To increase the signal over noise ratio, we apply the improved multihit and extended smearing techniques. We detail the gauge invariant squared components of the colour electric and colour magnetic fields both in the  mediator plane between the static quark and static antiquark and in the planes of the sources. We fit the field densities with appropriate ansatze and we observe the screening of the colour fields in all studied planes together with the quantum widening of the flux tube in the mediator plane. 
All components squared of the colour fields are non-vanishing and are consistent with a penetration length $\lambda \sim 0.22$ to 0.24 fm and an effective screening mass $\mu \sim 0.8 $ to 0.9 GeV. The quantum widening of the flux tube is well fitted with a logarithmic law in $R$.
\end{abstract}

\maketitle

PACS{11.15.Ha, 12.38.Gc,74.25.Uv,11.25.-w}

\section{Introduction}

Confinement in QCD remains a central problem of strong interactions. It has already been established, both from gauge invariant lattice QCD simulations
\cite{DiGiacomo:1989yp,DiGiacomo:1990hc,Singh:1993jj,Bali:1994de}
 and from experimental observations like Regge trajectories
 \cite{Regge:1959mz,Regge:1960zc,Collins:1977jy,Kaidalov:2001db,Bugg:2004xu},  
that the quark-antiquark confining potential is linear, and that a flux tube develops between quark-antiquark static charges. Even in dynamical QCD where the flux tube breaks due to the creation of another quark and antiquark, a flux tube develops up to moderate quark-antiquark distances. 
Recently, the flux tubes have been shown to also occur in lattice QCD simulations of different exotic hadrons  
\cite{Cardoso:2009kz,Cardoso:2011fq,Cardoso:2011cs,Cardoso:2012uk,Cardoso:2012rb}.
Here we return to the fundamental quark-antiquark flux tube, to measure in detail the profile of the SU(3) pure gauge lattice QCD flux tube in dimensions $D=3+1$. We parametrize the flux tube profile, providing new data for a better understanding of the confinement in QCD.

\begin{figure}[t!]
\begin{center}
    \includegraphics[width=9cm]{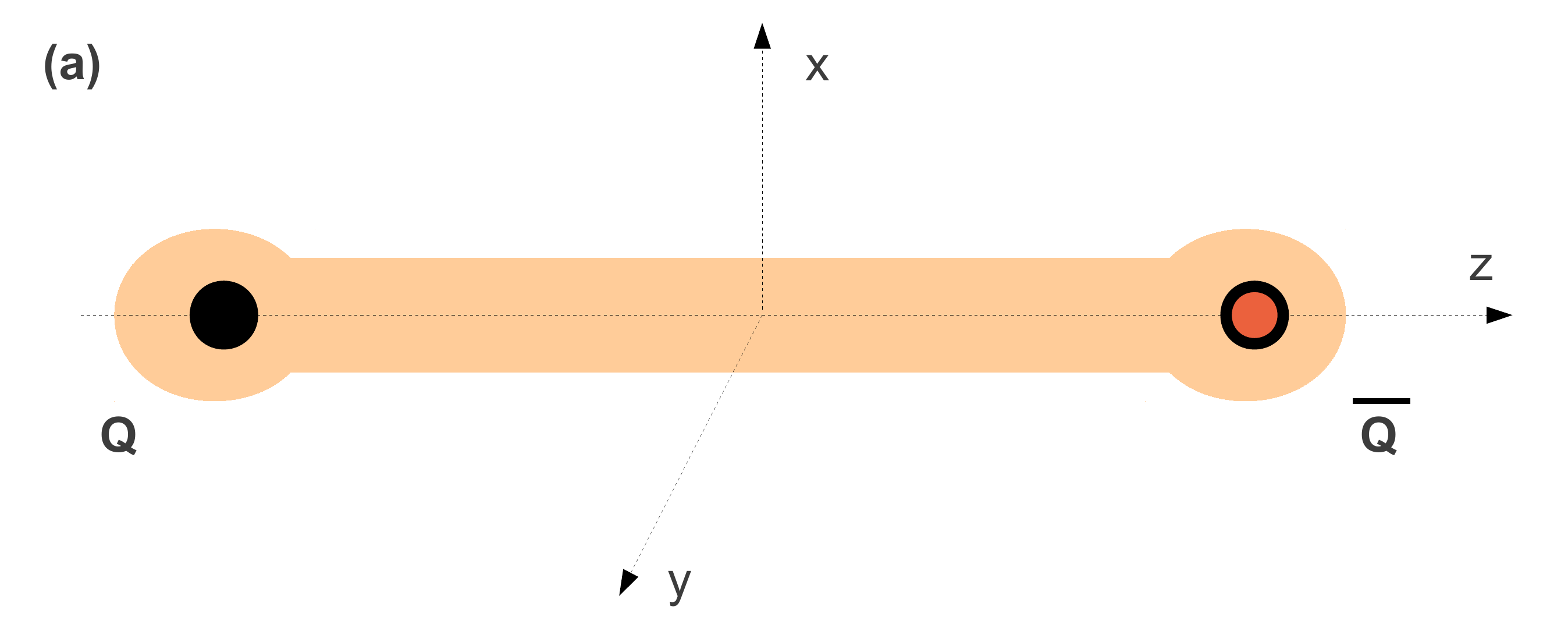}
\\
\vspace{.5cm}
    \includegraphics[width=9cm]{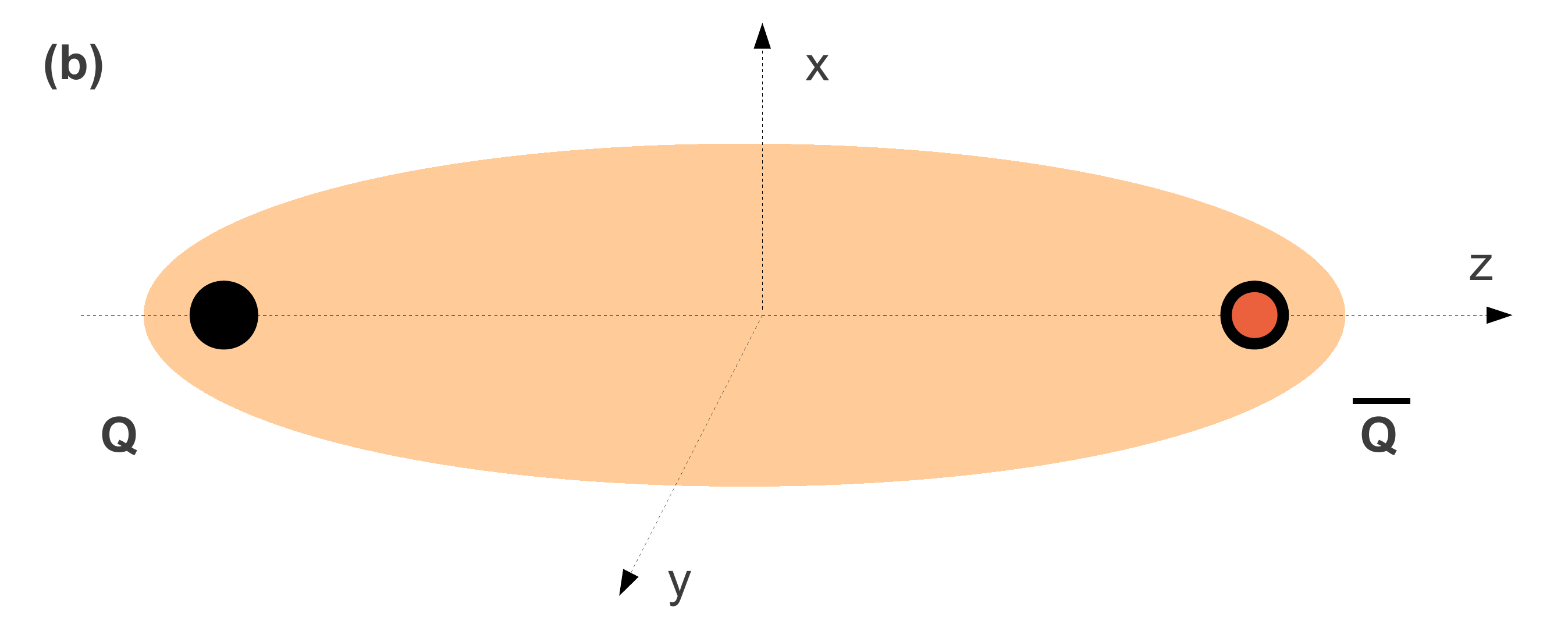}
    \caption{In (a) we illustrate a classical flux tube, similar to a solution of the Ginzburg-Landau and Amp\`ere equations for a superconductor. In (b) we illustrate a quantum flux tube, as in a Lattice QCD simulation, where the widening of the flux tube occurs due to the zero mode string vibration. The squeezing of the flux tube due to the colour screening in (a) is masked by the widening in (b).}
    \label{fig:screening-widening}
\end{center}
\end{figure}

In particular, presently two different perspectives for the QCD flux tube exist, possibly leading to the two different flux tubes of Fig. \ref{fig:screening-widening}, and we quantitatively compare them.


Already in the 1970's,  Nambu \cite{Nambu:1974zg}, 't Hooft \cite{'tHooft:1979uj} and Mandelstam \cite{Mandelstam:1974pi} proposed that quark confinement would be physically interpreted using the dual version of the superconductivity \cite{Baker:1989qp,Baker:1991bc}. The QCD vacuum state would behave like an ordinary magnetic superconductor, where Cooper-pair condensation leads to the Meissner effect, and
the magnetic flux is excluded or squeezed in a quasi-one-dimensional tube, the Abrikosov vortex, where the magnetic flux is quantized topologically. Magnetic charges are confined by Abrikosov-Nielsen-Olesen (ANO) vortices \cite{Abrikosov:1956sx,Nielsen:1973cs,Cardoso:2006mf} in an ordinary superconductor (Meissner effect).
Thus, it is important for the understanding of confinement in QCD to measure the flux tube profile, and to parametrize the colour screening \cite{Polyakov:1975rs,Banks:1977cc,Smit:1989vg,Bali:1996dm,Gubarev:1999yp,Koma:2003gq,Chernodub:2005gz}. 
Moreover the penetration length can be related as
\be
\lambda=  \mu^{-1}
\ee
to a possible effective mass $\mu$ of the dual gluon, if we further explore the analogy between QCD and superconductors where the field in the London equation has a direct relation with an effective mass of the interaction particle fields, i. e., the photon.
The dual gluon mass has been studied by several authors, \cite{Burdanov:1998tf,Jia:2005sp,Suzuki:2004uz,Suganuma:2004gq,Suganuma:2004ij,Suganuma:2003ds,Kumar:2004fj,Burdanov:2002ne}, as well as the gluon effective mass, see Ref. \cite{Cardoso:2010kw} for a review of  the dual gluon and gluon effective masses present in the literature. Interestingly, there is also an evidence for a gluon mass in the Landau Gauge 
\cite{Oliveira:2010xc} 
and in the multiplicity of particles produced in heavy ion collisions
\cite{Bicudo:2012wt}. 
Recently  the penetration length started to be computed with gauge invariant lattice QCD techniques 
\cite{Cardoso:2010kw,Cardaci:2010tb,Cea:2012qw}. 
In superconductors another parameter, the coherence length $\xi$ is defined as well and related to the curvature of the flux tube profile.

\begin{figure}[t!]
\begin{centering}
 \includegraphics[width=1.0\columnwidth]{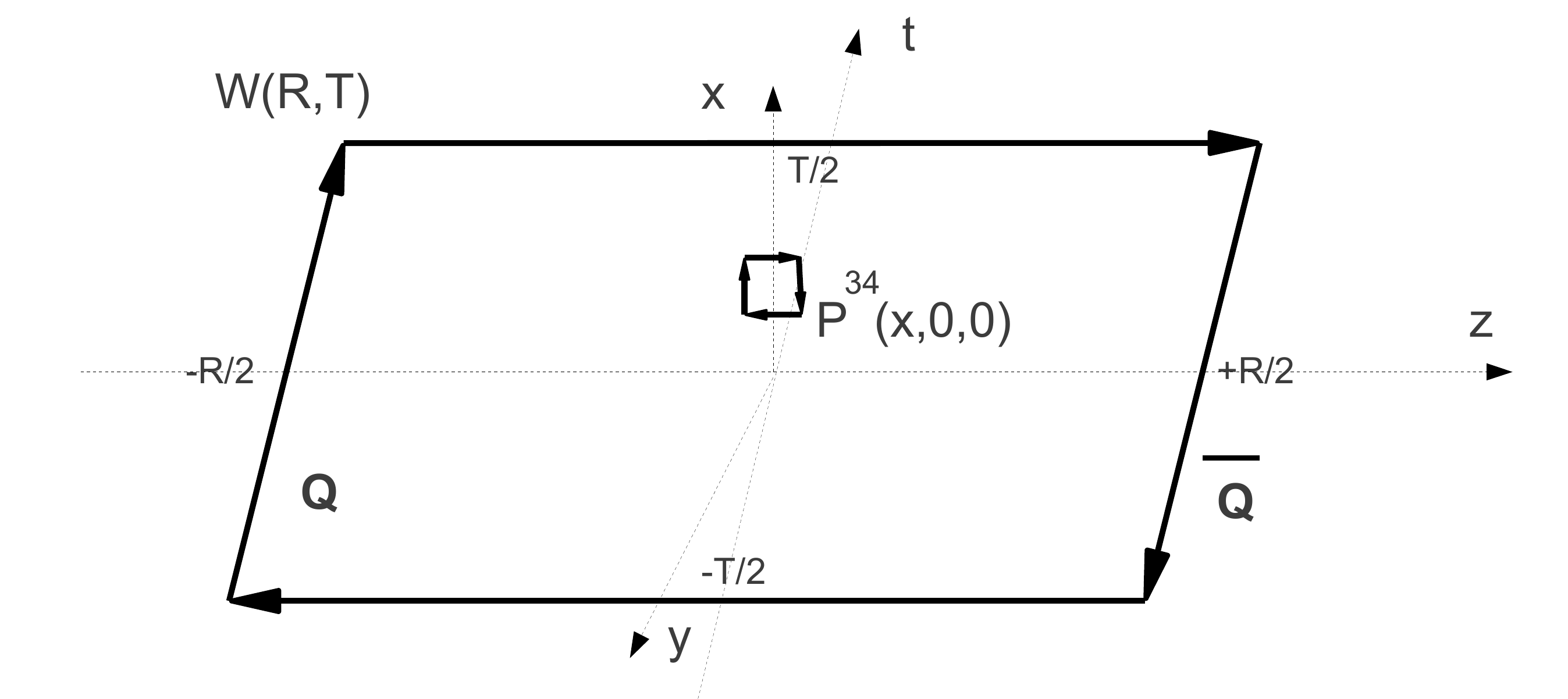}\tabularnewline
\par\end{centering}
\caption{Wilson loop and example of a plaquette for the computation of the electric field squared, where we project the $D=3+1$ space time  in a plane including the $z$ axis. We compute the fields squared in three planes perpendicular to the charge-anticharge axis: in the mediator plane of the charges as illustrated in this Figure, in the plane of the charge and in the plane of the anti-charge.
\label{loop_plaquette}}
\end{figure}


On the other hand, at quark-antiquark distances larger than the penetration length, the flux tube is similar to a quantum string. And the quantum string vibrates, even in the groundstate where it has zero mode vibrations. 
A fair description of the fundamental QCD flux tube - with charges in the triplet representation of SU(3) - is given by the string
model, based on the \foreignlanguage{american}{Nambu-Goto} Action
\cite{Nambu:1978bd,Goto:1971ce},
\be
S=-\sigma\int d^{2}\Sigma \ .
\ee
The energy of the quantum string with length $R$ and fixed ends, with quantum transverse fluctuations quantum number $n$, is expressed in the L\"uscher term  and in the Arvis Potential \cite{Luscher:1980iy,Arvis:1983fp},
\bea
V_{n}(R)&=&\sigma\sqrt{R^{2}+\frac{2\pi}{\sigma}(n-\frac{D-2}{24})}
\non \\
&=&\sigma R+\frac{\pi}{R}(n-\frac{D-2}{24})+\ldots
\label{Arvis}
\eea
In Eq. (\ref{Arvis}), $D$ is the dimension of the space time. Note that the Arvis potential is tachionic at small distances since the argument of the square root is negative, moreover rotational invariance is only achieved for $D=26$.
Nevertheless  the first two terms in the $1/R$ expansion are more general that the Arvis potential, since they fit the $D=3$ and $D=4$ lattice data quite well beyond the tachionic distance. 
The Coulomb term is independent of the string tension $\sigma$ and for the physical
$D=3+1$ has the value $-\frac{\pi}{12}$. This is the L\"uscher term  \cite{Luscher:1980iy}.
The energy spectrum of a static quark-antiquark and of its flux tube is certainly well defined (not tachionic) and this was the first evidence of flux tube vibrations found in lattice field theory. Moreover it was shown \cite{Luscher:1980iy} that the width of the groundstate  flux tube  diverges when $R\rightarrow\infty$ with a logarithmic law,
\begin{equation}
w^{2}\sim w_{0}^{2}\,\log\frac{R}{R_{0}}
\end{equation}
where $w^{2}$ is the mean squared radius of the flux tube. 
This enhancement of the the flux tube transverse radius as $R\rightarrow\infty$ is called
widening. The widening as been recently extended with two-loop calculations \cite{Gliozzi:2010zt}.
So far widening has been verified numerically 
for compact U(1) QED $D=2+1$ lattices \cite{Amado:2012wt} and for non-abelian SU(2) $D=2+1$ lattices \cite{Armoni:2008sy,Armoni:2008sy,Bakry:2010zt,Bakry:2010sp,Bornyakov:2002vt,Bornyakov:2003gn,Cardoso:2012aj,Caselle:1995fh,Caselle:2010zs,Caselle:2012rp,Chernodub:2007wi,Giudice:2006xe,Giudice:2006hw,Gliozzi:2006sj,Gliozzi:1994bc,Gliozzi:2010jh,Gliozzi:2010zv,Greensite:2000cs,Lucini:2001nv,Meyer:2010tw}. The widening in SU(3) lattice QCD and in $D=3+1$, which is the pure gauge closer to real strong interactions, has not been measured previously.


In this paper, we present a SU(3) gauge independent lattice QCD computation in $D=3+1$ for the penetration length and of the string quantum widening. We think this is a premi\`ere both for the study of widening in SU(3) and for dimension as large as $D=3+1$. This is also the first attempt to separate the screening from the quantum widening. While the screening leads to an exponential decay of of the flux tube profile, the widening leads to a gaussian profile.

In section II, we introduce the lattice QCD formulation.
We briefly review the Wilson loop for this system, which was used in Bicudo et al. \cite{Bicudo:2007xp},  Cardoso et al. \cite{Cardoso:2007dc} and Cardoso et al. \cite{Cardoso:2009kz}, and show how we compute the colour fields and as well as the lagrangian and energy densities distributions.
In Section III we show the techniques we utilize to increase the signal over noise ratio.
In Section IV we discuss our ansatz for the the width of the QCD flux tube.
In section V, the lattice numerical results are shown together with their fits.
Finally, we present the conclusion in section VI.

\section{Computation of the Chromo-fields in the flux tube}

\begin{figure}[t!]
\begin{centering}
\begin{tabular}[t]{cc}
\includegraphics[width=0.5\columnwidth]{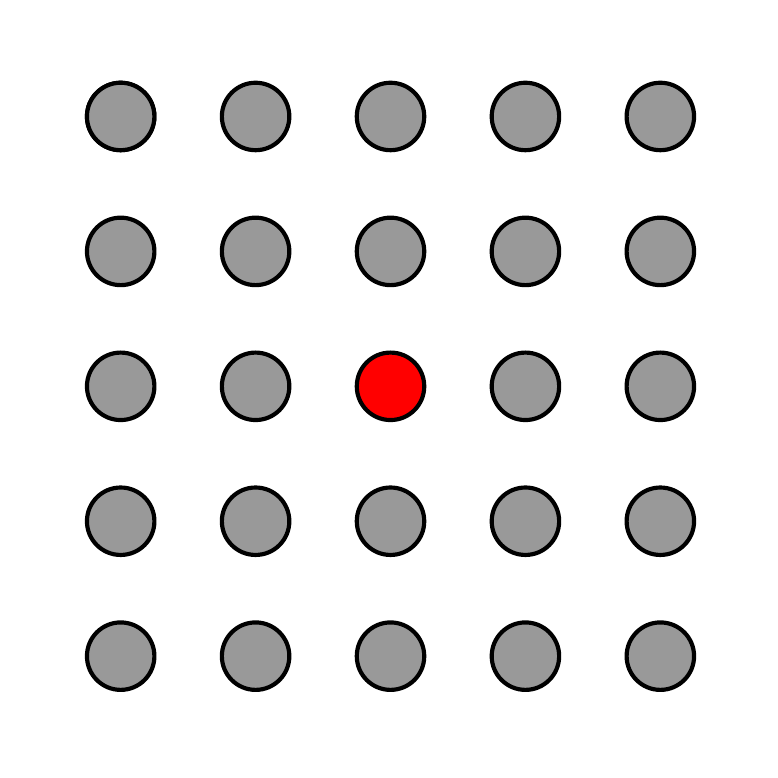} & \includegraphics[width=0.5\columnwidth]{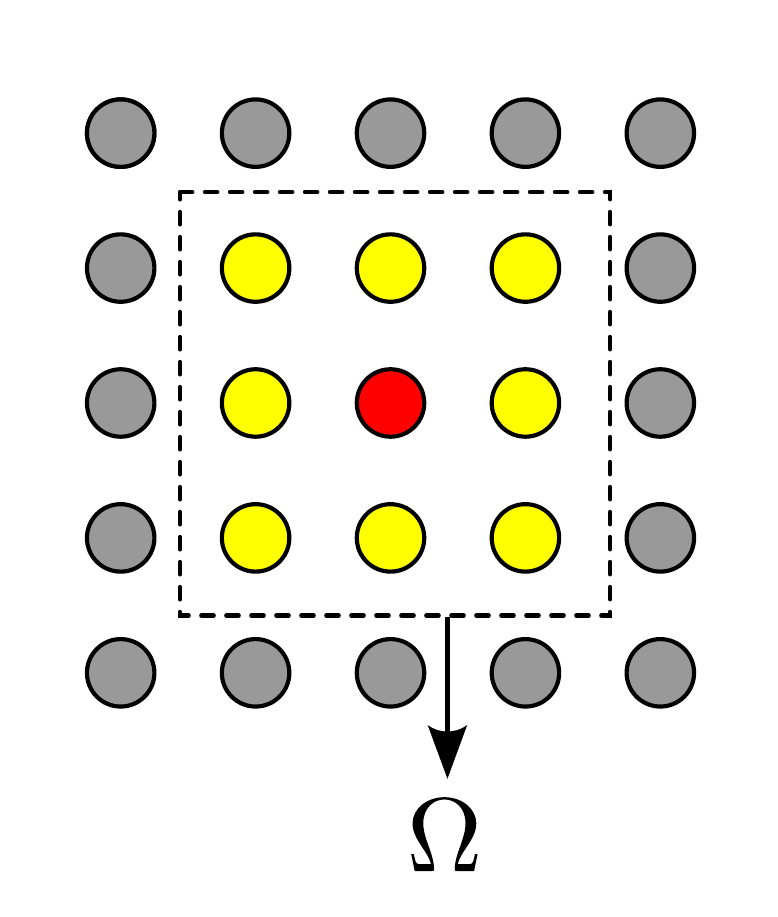}\tabularnewline
\end{tabular}
\par\end{centering}
\caption{Left: Simple Multihit. Right: Extended Multihit.
\label{fig:multihit}}
\end{figure}

We impose our static quark-antiquark system with the standard Wilson $W(R,T)$ loop 
\cite{Wilson:1974sk},
\bea
W(R,T)&=&\mbox{Tr} \biggl[ \, U_{\mu}(0,0,{-R \over 2},{-T \over 2})\ldots U_{\mu}(0,0,{-R \over 2}-1,{-T \over 2})
\non \\
&&  U_{4}(0,0,{R \over 2},{-T \over 2})\ldots U_{4}(0,0,{R \over 2},{T \over 2}-1)
\non \\
&& U_{\mu}^{\dagger}(0,0,{R \over 2}-1,{T \over 2})\ldots U_{\mu}^{\dagger}(0,0,{-R \over 2},{T \over 2})
\non \\
&& U_{4}^{\dagger}(0,0,{-R \over 2},{T \over 2}-1)\ldots U_{4}^{\dagger}(0,0,{-R \over 2},{T \over 2})\, \biggr] \ .
\eea
In the limit of large euclidean time limit $T\rightarrow\infty$, the expectation value
\be
\langle W(R,T)\rangle=\sum_{n}|C_{n}|^{2}e^{-V_{n}T} 
\ee
selects the groundstate of the static quark-antiquark system, aligned in the $z$ direction with an intercharge distance $R$.

To compute the gauge invariant squared components of the chromoelectric and chromomagnetic
fields on the lattice, we utilize the Wilson loop and plaquette $P_{\mu\nu}$ expectation values,
\bea
    \Braket{{B_i}^2(\mbf r)} &=& \frac{\Braket{W(R,T)\,P(\mbf r)_{jk}}}{\Braket{W(R,T)}}-\Braket{P(\mbf r)_{jk}} \, ,
\non 
\\
   \Braket{{E_i}^2(\mbf r)} &=& \Braket{P(\mbf r)_{0i}}-\frac{\Braket{W(R,T) \,P(\mbf r)_{0i}}}{\Braket{W(R,T)}} \, ,
\label{fields} 
\eea
where the $jk$ indices of the plaquette complement the index $i$ of the magnetic field.
The plaquette at position $\mbf r=(x,y,z)$ is computed at lattice euclidian time $t=0$, as depicted in Fig. \ref{loop_plaquette}.
In Eq. (\ref{fields}) we subtract, from the plaquette computed in the presence of the static charges, the average plaquette computed in the vacuum. This cancels the  vacuum fluctuations of the fields. To get the plaquette in the lattice vertices, we average the neighbouring plaquettes.


\begin{figure}
\begin{centering}
\begin{tabular}{ccc}
\includegraphics[bb=130bp 100bp 280bp 280bp,clip,width=0.15\textwidth]{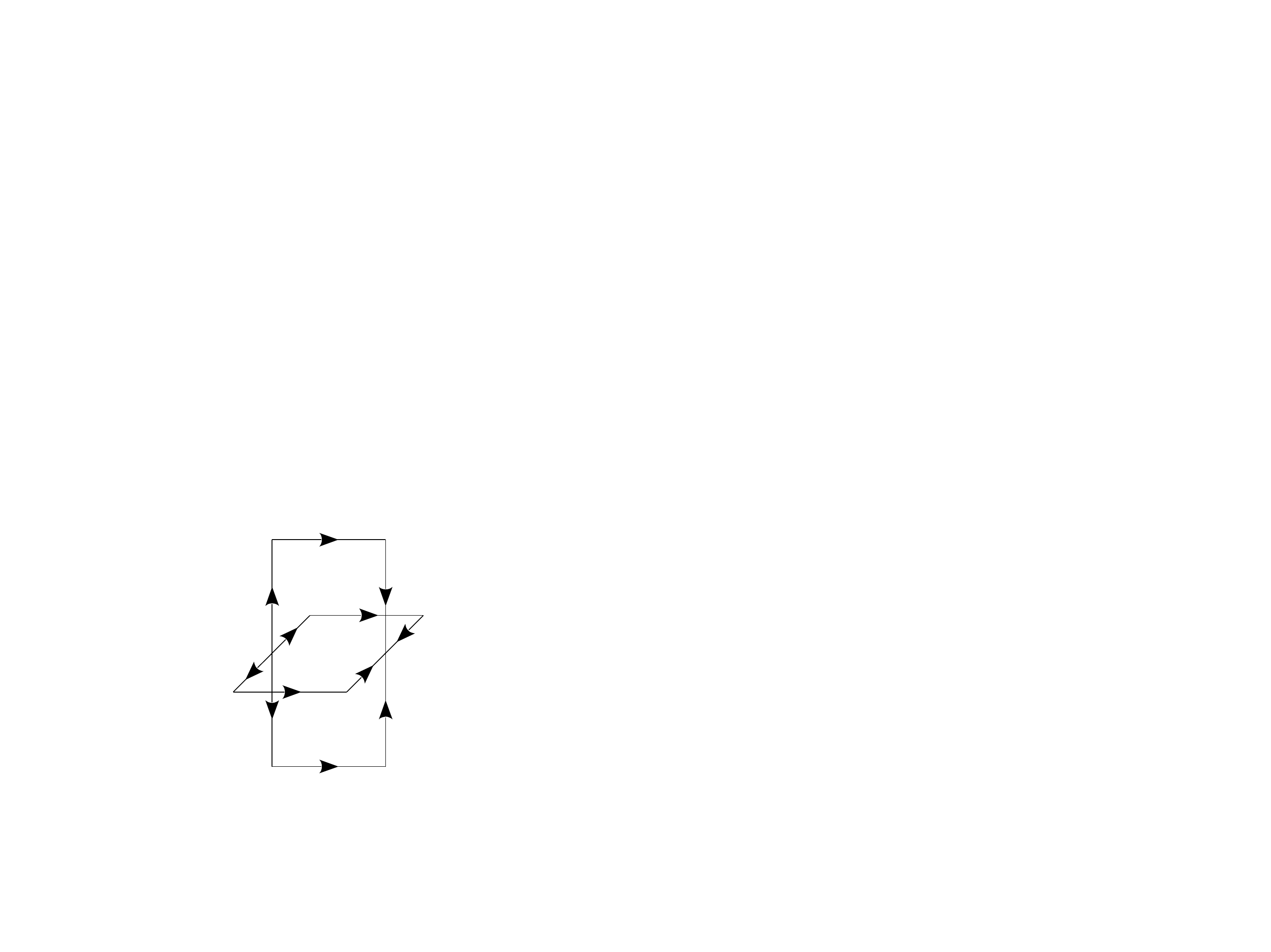} &  
\includegraphics[bb=320bp 80bp 480bp 300bp,clip,width=0.165\textwidth]{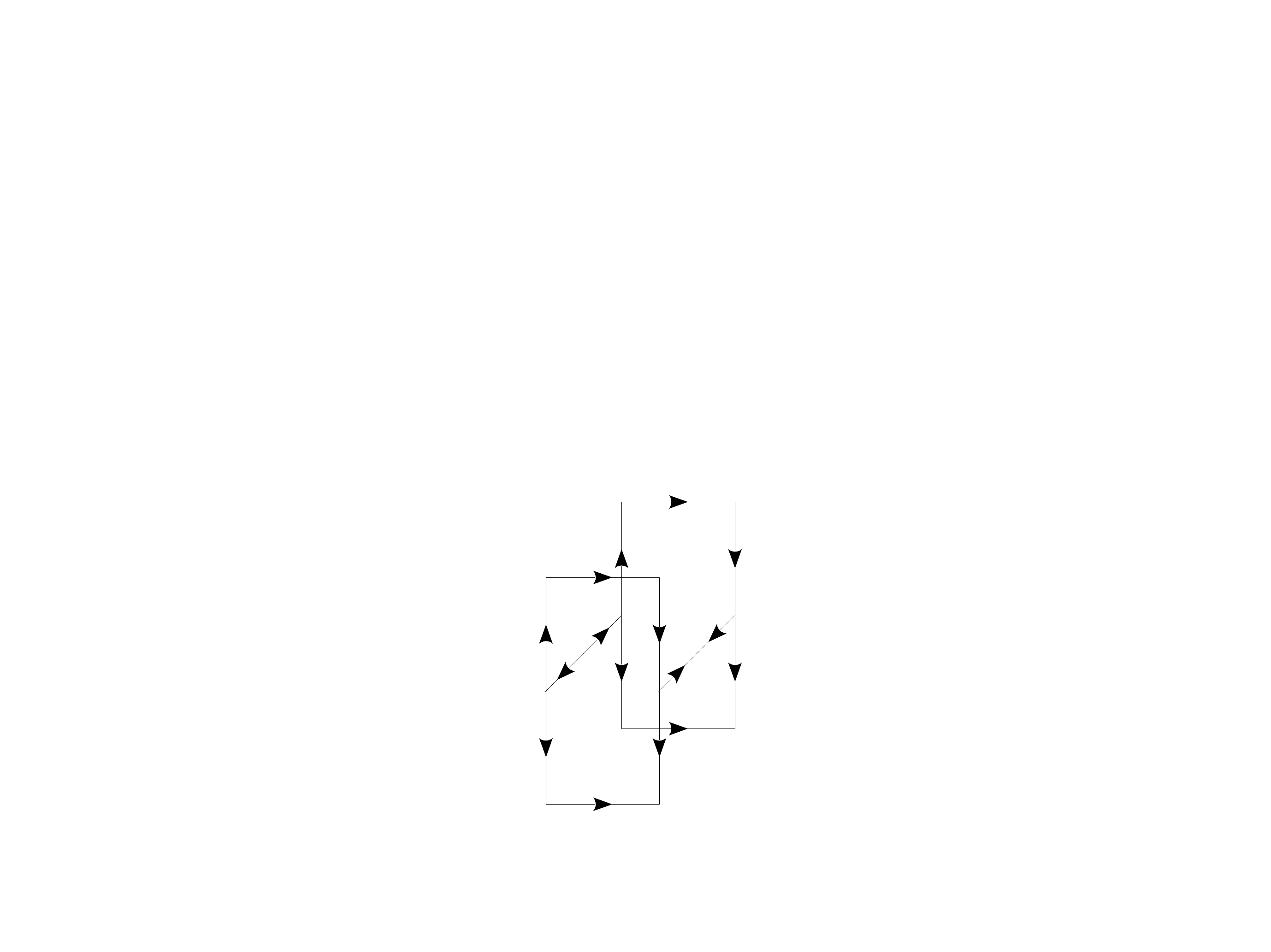} & 
\includegraphics[bb=510bp 30bp 700bp 340bp,clip,width=0.195\textwidth]{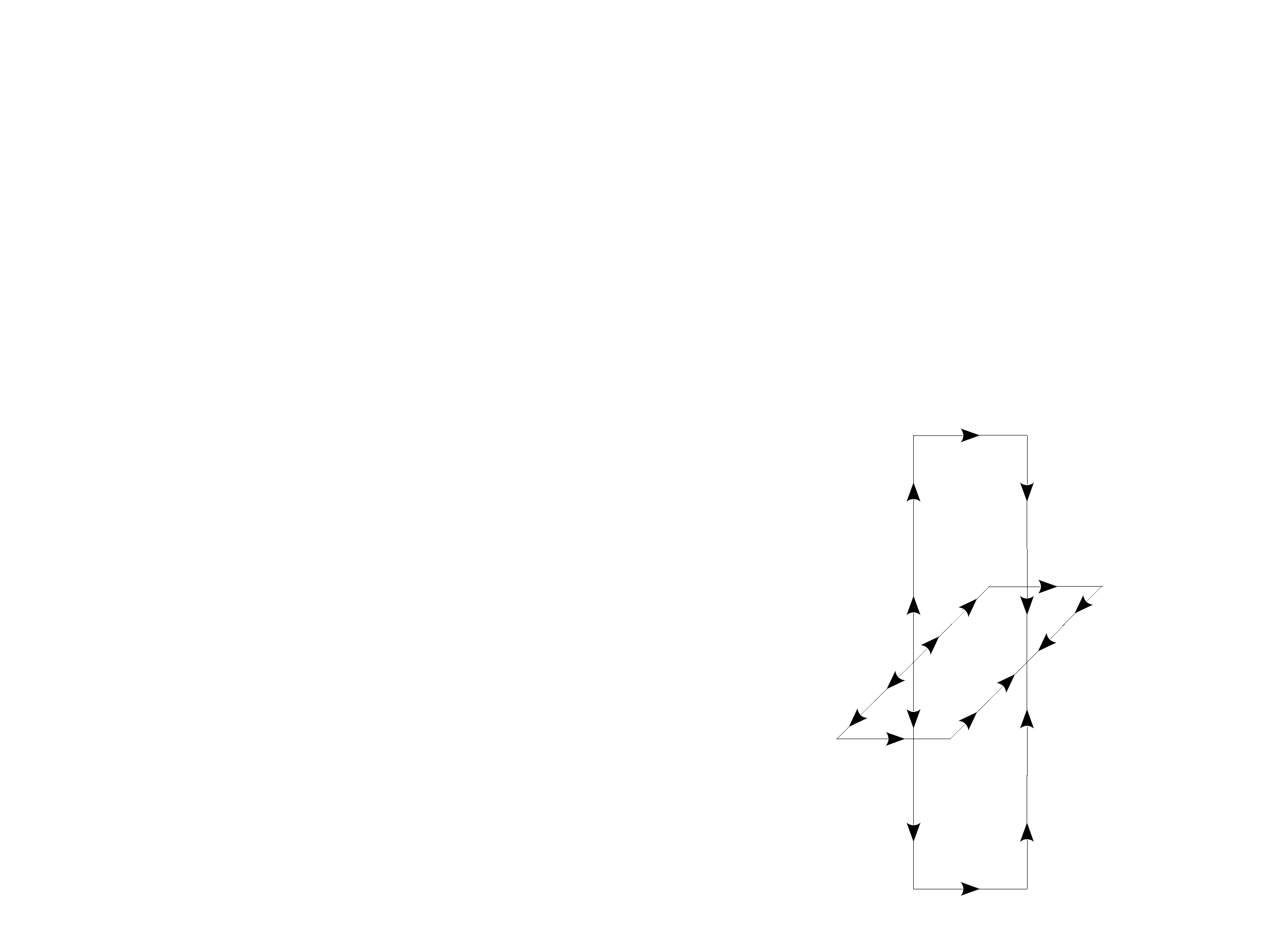}\tabularnewline
\end{tabular}
\par\end{centering}
\caption{Staples used in the extended spatial smearing. \label{fig:staples}}
\end{figure}

We define our plaquette as,
\be
P_{\mu\nu}\left(\mbf r \right)=1 - \frac{1}{3} \ReC\,\Tr\left[ U_{\mu}(\mbf r) U_{\nu}(\mbf r+\mu) U_{\mu}^\dagger(\mbf r+\nu) U_{\nu}^\dagger(\mbf r) \right]\ ,
\label{plaquettewithtrace}
\ee
which, for small lattice spacing $a$, can be expanded in a series of powers of the symmetric tensor $F^{\mu \nu \ c}$, which components are the electric and magnetic field components. Prior to performing the trace, the expansion reads \cite{Creutz:1984mg,Gattringer:2010zz},
\bea
P_{\mu\nu}  &=&  1 - {1 \over 3} \ReC \,  \Tr \exp  \left[ i  g a^2 \sum_c   F_{\mu\nu} ^c  T^c + {\cal O} (a^3) \right]
\non \\ 
&=&   \ReC \, \Tr \biggl\{ {1 \over 36} g^2 a^4 \left[ F_{\mu\nu} ^c F_{\mu\nu} ^c   + {\cal O}(a) \right] I
\non \\
&& \ \ \ \ \ \ \ \ \ \   -  { i \over 3} g a^2 \sum_c  \left[ F_{\mu\nu} ^c + {\cal O}(a) \right] T^c   \biggr\}
\label{notrace}
\eea
where $T^c = \lambda^c / 2$ are the generators of the Lie algebra and $I$ is the identity matrix.
In abelian theories, such as U(1) QED, the electric and magnetic fields components can be computed with the plaquette at order $a^2$ and are gauge invariant. In non-abelian gauge theories, such as SU(3), the electric and magnetic field components are not gauge invariant since they depend on the colour index $c$. 
In SU(3) we have to go up to order $a^4$ to find our first non-vanishing gauge invariant term in the plaquette expansion, and it is the square of a component of the electric or magnetic fields. For instance ${E_x}^2= \sum_c ( {E_x}^c)^2 $ is gauge invariant, while ${E_x}^c$ is not. Thus, to directly produce the squared components, we perform the trace.

Notice the field densities defined in Eq. (\ref{fields}) are dimensionless. To arrive at physical units , 
\be
\sum_c F_{\mu\nu} ^c  F_{\mu\nu} ^c  = {2 \beta  \over a^4 } \left[ 1 - { 1 \over 3} Tr\left(P_{\mu\nu} \right) \right] +  {\cal O}(a)
\ee
we have to multiply the dimensionless field densities by $2 \beta / a^4$.

The classical energy ($\mathcal{H}$) and the lagrangian ($\mathcal{L}$) densities are directly computed from the filed densities,
\bea
   \Braket{ \mathcal{H}(\mbf r) } &=& \frac{1}{2}\left( \Braket{\mbf E^2(\mbf r)} + \Braket{\mbf B^2(\mbf r)}\right)\ , \\
    \label{energy_density}
   \Braket{ \mathcal{L}(\mbf r) } &=& \frac{1}{2}\left( \Braket{\mbf E^2(\mbf r)} - \Braket{\mbf B^2(\mbf r)}\right)\ ,
    \label{lagrangian_density}
\eea
and we can utilize any of the densities, either of the squared component of the fields, of the action or of the classical energy, to study the profiles of the flux tubes.

\begin{figure}[t!]
\begin{centering}
\includegraphics[width=0.45\textwidth]{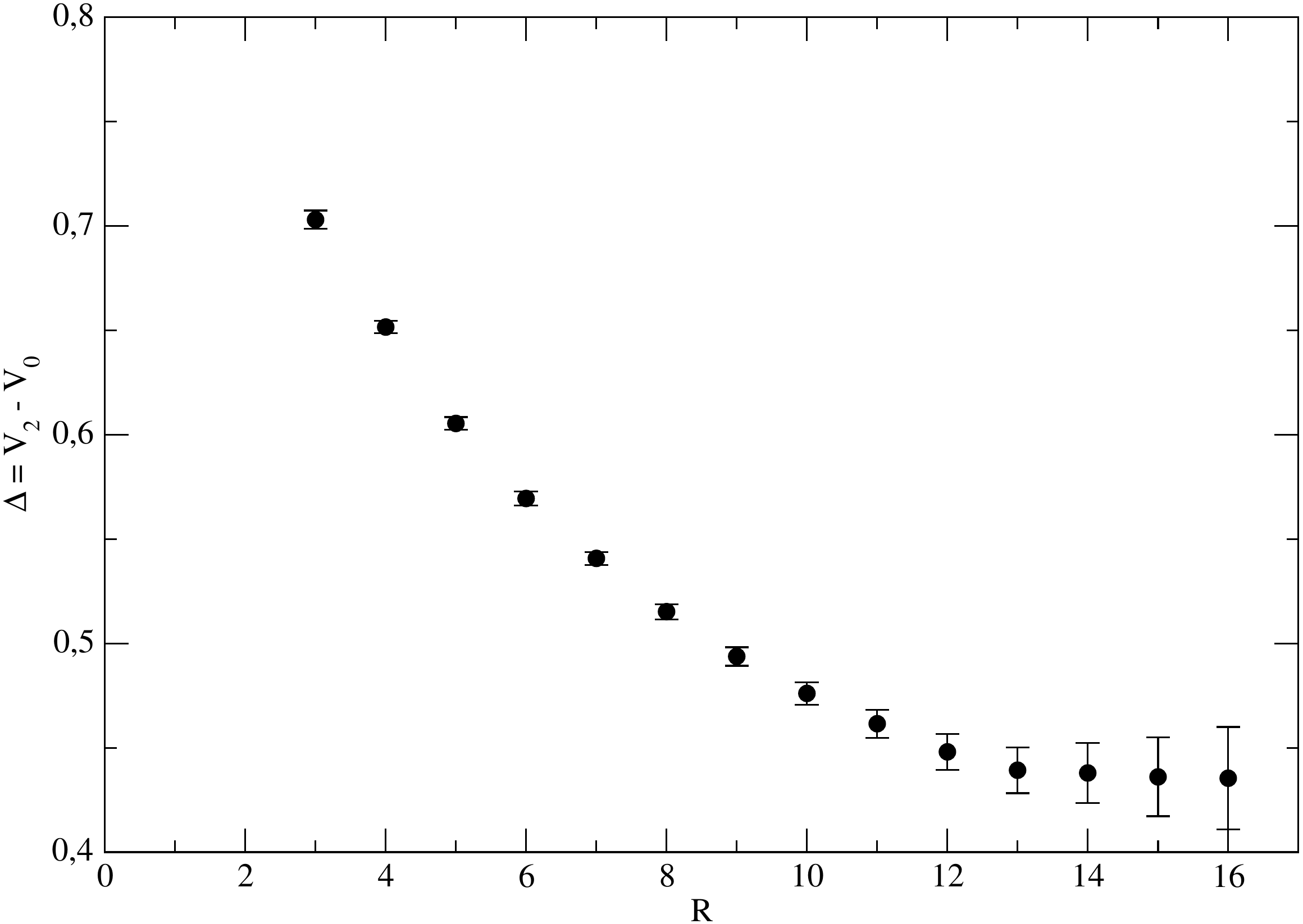}
\par\end{centering}
\caption{Gap between the first excited state and the ground state $\Delta=V_{2}-V_{1}$
as a function of $R$.}
\label{fig:excitedgap}
\end{figure}

\section{Techniques employed to improve the signal}

To compute the static field expectation value, we plot the expectation value
    $ \Braket{E^2_i(\mbf r)} $ or   $\Braket{B^2_i(\mbf r)}$ as a function of the temporal extent $T$ of
    the Wilson loop. At sufficiently large $T$, the groundstate corresponding to the
studied quantum numbers dominates, and the expectation value tends to a horizontal plateau. 
To compute the fields, we fit the horizontal plateaux obtained for each point $\mbf r$ 
 For the distances $R$ considered, we find in the range of $T\in [4,12]$ in lattice units,
horizontal plateaux with a $\chi^2$ /dof $\in [0.3,2.0] $.
We finally compute the error bars of the fields with the jackknife method. 

To produce the expectation values, we utilize 1100 pure gauge $32^{4}$ configurations with $\beta=6.0$. This beta corresponds to the lattice spacing
$a=0.0983737$ fm and  $a^{-1}=2.00257T$ GeV
\cite{Edwards:1997xf}.

\begin{figure}[t!]
\begin{centering}
\includegraphics[width=0.45\textwidth]{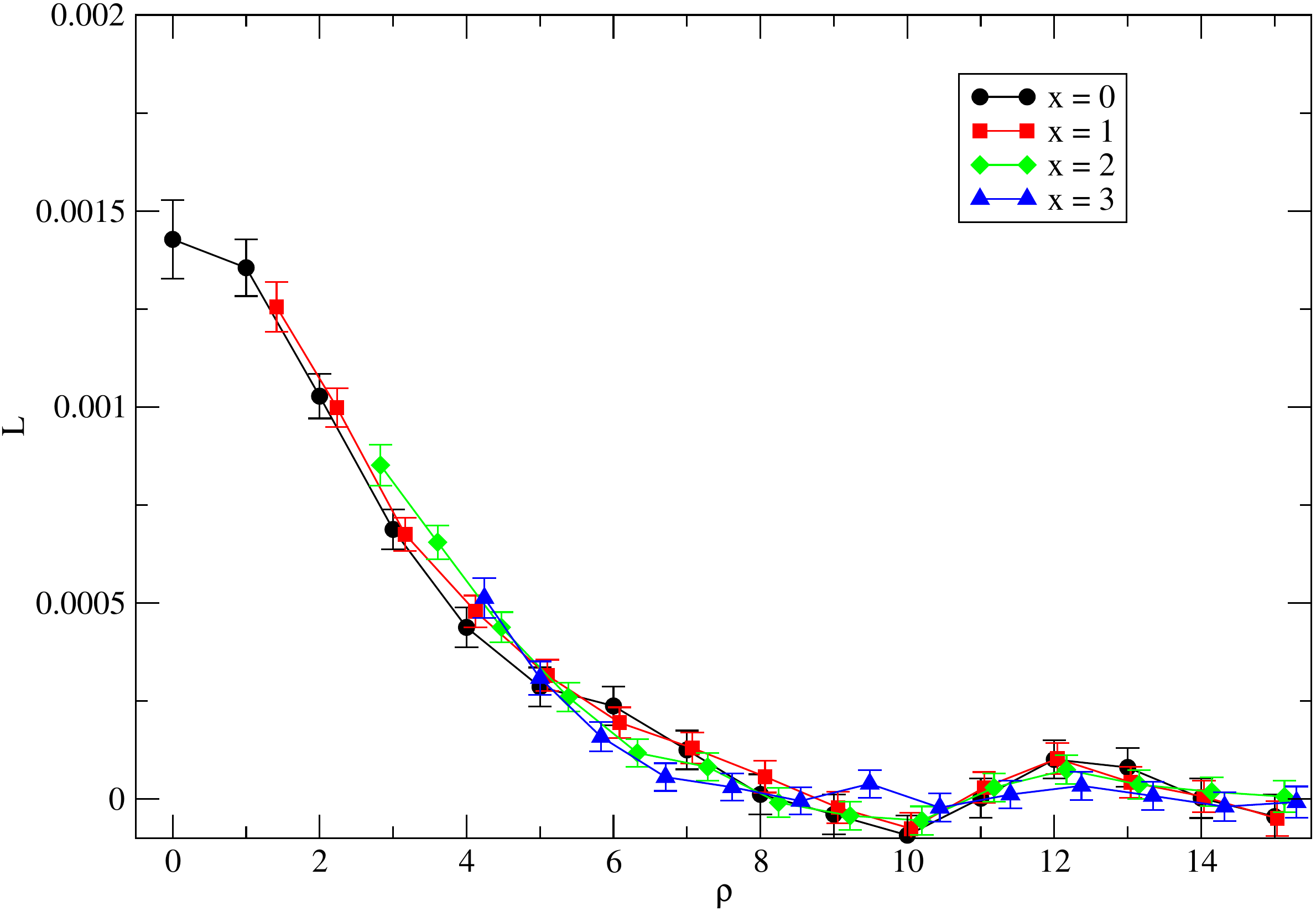}
\par\end{centering}
\caption{Lagrangian density computed in the mediator plane for inter-charge distance of $R=12$, as a function of the cylindrical distance $r$. We plot separately the density measured in different lines of the mediator plane with fixed $x$. At large distances $r$, the lattice artefacts, due to the square and finite lattice, produce systematic errors already larger than the statistical error bars represented in the Figure. This shows that our statistical noise are sufficient reduced by the extended multihit, the extended spatial smearing and the variational basis methods. }
\label{fig:artifacts}
\end{figure}

In order to reduce the noise, we utilize an improved version of the
multihit illustrated in Fig. \ref{fig:multihit} and an extended spatial smearing technique with staples shown in Fig. \ref{fig:staples}. Moreover, to
reduce the contamination of the groundstate from excited states, we
use the energy gap between the first excited and ground states, depicted in Fig. \ref{fig:excitedgap}
calculated using a variational basis. With all three combined techniques, we are able to get a clear signal, with statistical errors already smaller than the lattice artefacts plotted in Fig. \ref{fig:artifacts}.

\begin{figure*}[t!]
\begin{center}
 \includegraphics[width=1.0\textwidth]{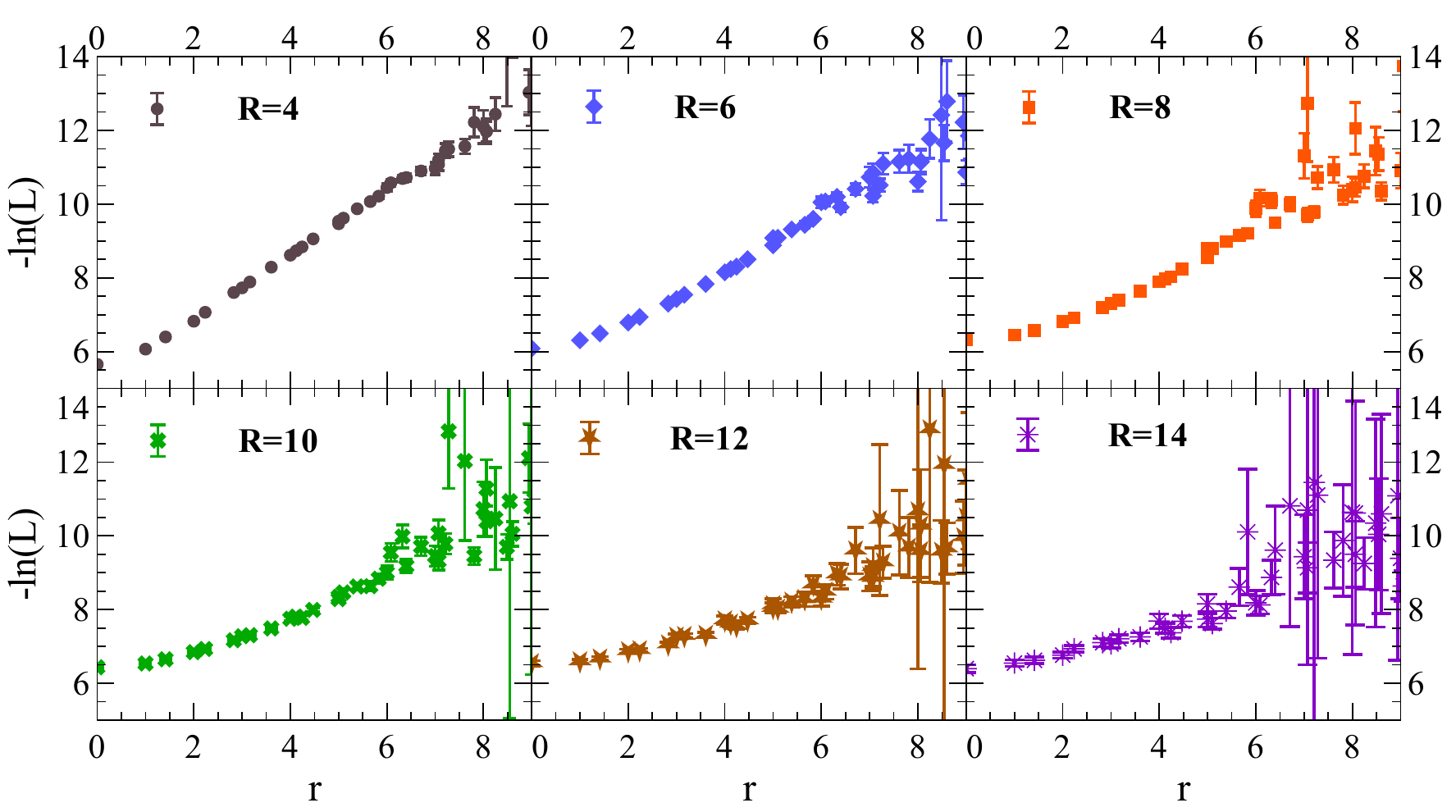}
     \caption{
Results for minus the log of the action density in the charge mediator plane, for $R=4, \ R=6, \ R=8, \ R=10, \ R=12,$ and $ R=14$. The plots suggest the exponent is quadratic at small distances and linear at large distances, in agreement with our ansatz.
    \label{fig:exponent}
}
\end{center}
\end{figure*}

\subsection{Extended Multihit}

In the multihit  \cite{Brower:1981vt,Parisi:1983hm} method we replace each temporal link by its thermal
average, with it's first neighbours fixed, that is $U_{4}\rightarrow\overline{U}_{4}=\frac{\int dU_{4}\, U_{4}\, e^{\beta\mbox{Tr}[U_{4}F^{\dagger}]}}{\int dU_{4}\, e^{\beta\mbox{Tr}[U_{4}F^{\dagger}]}}$.

We generalize this method by instead replacing each temporal link
by it's thermal average with the first $N$ neighbours fixed, that
is,
\begin{equation}
U_{4}\rightarrow\overline{U}_{4}=\frac{\int[\mathcal{D}U]_{\Omega}\, U_{4}\, e^{\beta\sum_{\mu\mathbf{s}}\mbox{Tr}[U_{\mu}(\mathbf{s})F_{\mu}^{\dagger}(\mathbf{\mathbf{s}})]}}{\int[\mathcal{D}U]_{\Omega}\, e^{\beta\sum_{\mu\mathbf{s}}\mbox{Tr}[U_{\mu}(\mathbf{s})F_{\mu}^{\dagger}]}}
\end{equation}

By using $N=2$ we are able to greatly improve the signal, when compared
with the error reduction achieved with the simple multihit. Of course,
this technique is more computer intensive than simple multihit, while
being simpler to implement than multilevel \cite{Luscher:2001up} and it's
application being independent in the value of $R$. The only restriction
is that $R\geq2N$ for this technique to be valid.

\subsection{Extended spatial smearing}

To increase the ground state overlap, we use a spatial extended APE, \cite{Falcioni:1984ei,Albanese:1987ds}
like smearing, namely

\begin{equation}
U_{i}\rightarrow\mathcal{P}_{SU(3)}\Big[U_{i}+w_{1}\sum_{j}S_{ij}^{1}+w_{2}\sum_{j}S_{ij}^{2}+w_{3}\sum_{j}S_{ij}^{3}\Big]
\end{equation}

the staples $S_{ij}^{1}$, $S_{ij}^{2}$ and $S_{ij}^{3}$ are the
ones shown in Fig.  \ref{fig:staples}. As can be seen, this technique
reduces to the common APE smearing when $w_{2}=w_{3}=0$.

\subsection{Variational basis to compute $\Delta$}

Even using this technique we were not able to find a value of $t$
for which the plaquette to Wilson Loop correlators are stable within
error bars, while still have a sufficiently high signal to noise ratio.
To solve this, we note that the correlator which gives the average
of field $\langle F\rangle$ should be given by the formula $\langle F\rangle_{t}=\langle F\rangle_{\infty}+b\, e^{-\Delta t}$
for large values of $t$, with $\Delta=V_{2}-V_{0}$, being the different
between the first excited state which has overlap with the Wilson
loop and the ground state potential. To compute $\Delta$, we a use
a variational basis 
\cite{Luscher:1990ck,Allton:1993wc}
of four levels of APE smearing, with the potentials
$V_{2}$ and $V_{0}$ being given by the solution of the variational
generalized eigensystem 
\begin{equation}
\langle W_{ij}(t)\rangle c_{j}^{n}(t)=w_{n}(t)\langle W_{ij}(0)\rangle c_{j}^{n}(t)\label{eq:variational}
\end{equation}
where $\langle W_{ij}\rangle=\langle\mathcal{O}_{i}(t)\mathcal{O}_{j}^{\dagger}(0)\rangle$
is the correlation between the meson creation and annihilation operators
at time $t$ and $0$ in the smeared states $i$ and $j$ respectively.

\begin{figure*}[t!]
\begin{center}
    \includegraphics[width=0.45\textwidth]{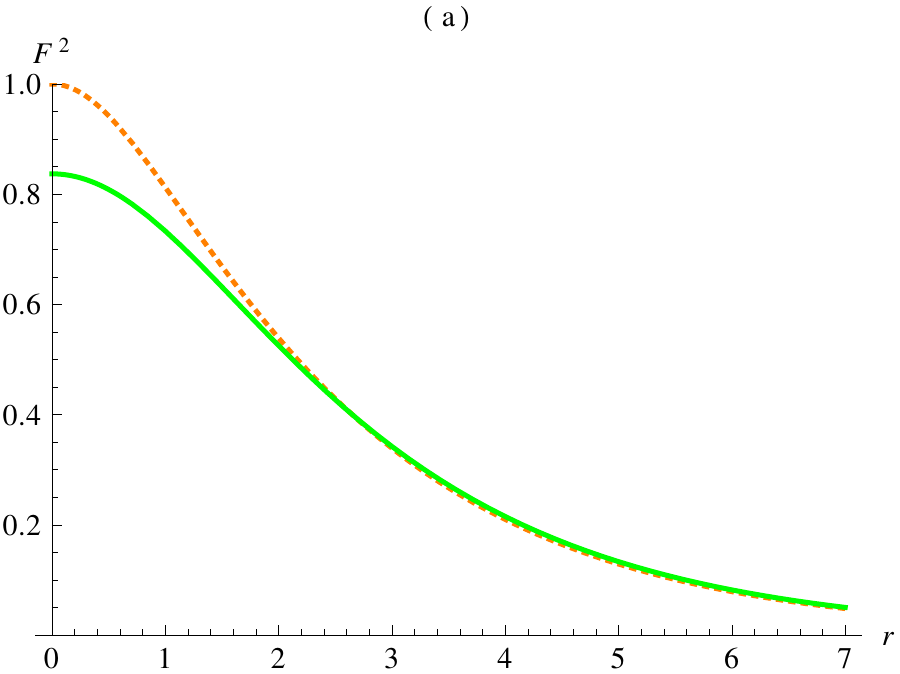}
\hspace{.5cm}
   \includegraphics[width=0.45\textwidth]{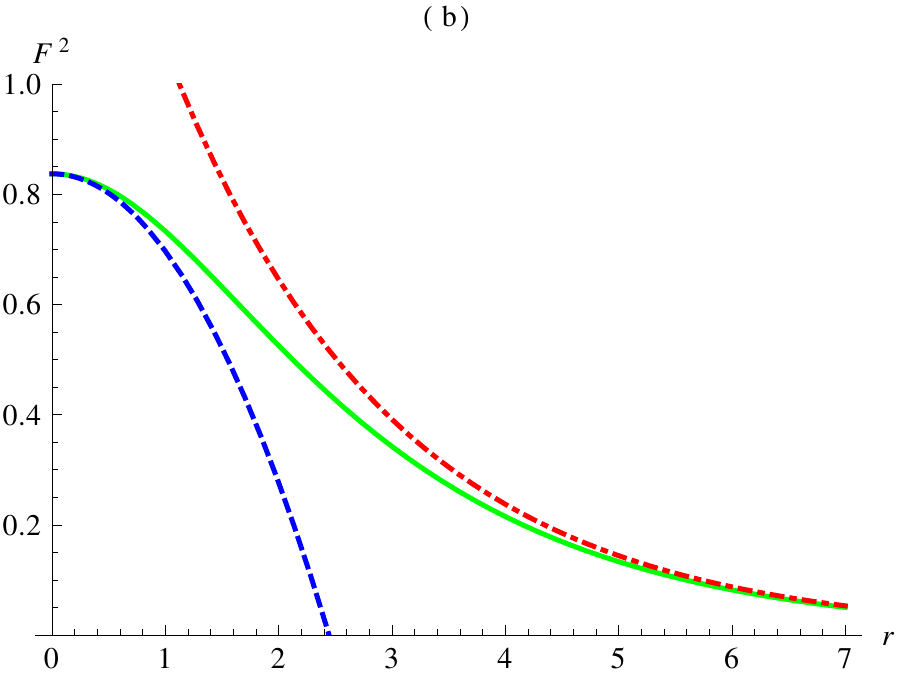}
     \caption{In (a) we the illustrate, for arbitrary parameters $ {E_0}^2=1$, $\lambda=1$, $\nu=1$, $\alpha=1$, our ansatz for the classical field and the quantum, or convoluted, one as a function of the distance to the charge axis $r$. In (b) the convoluted field and its large and small $r$ asymptotic functions are shown.}
    \label{fig:screening-convoluting}
\end{center}
\end{figure*}

\begin{figure*}[t!]
\begin{center}
  \includegraphics[width=1.0\textwidth]{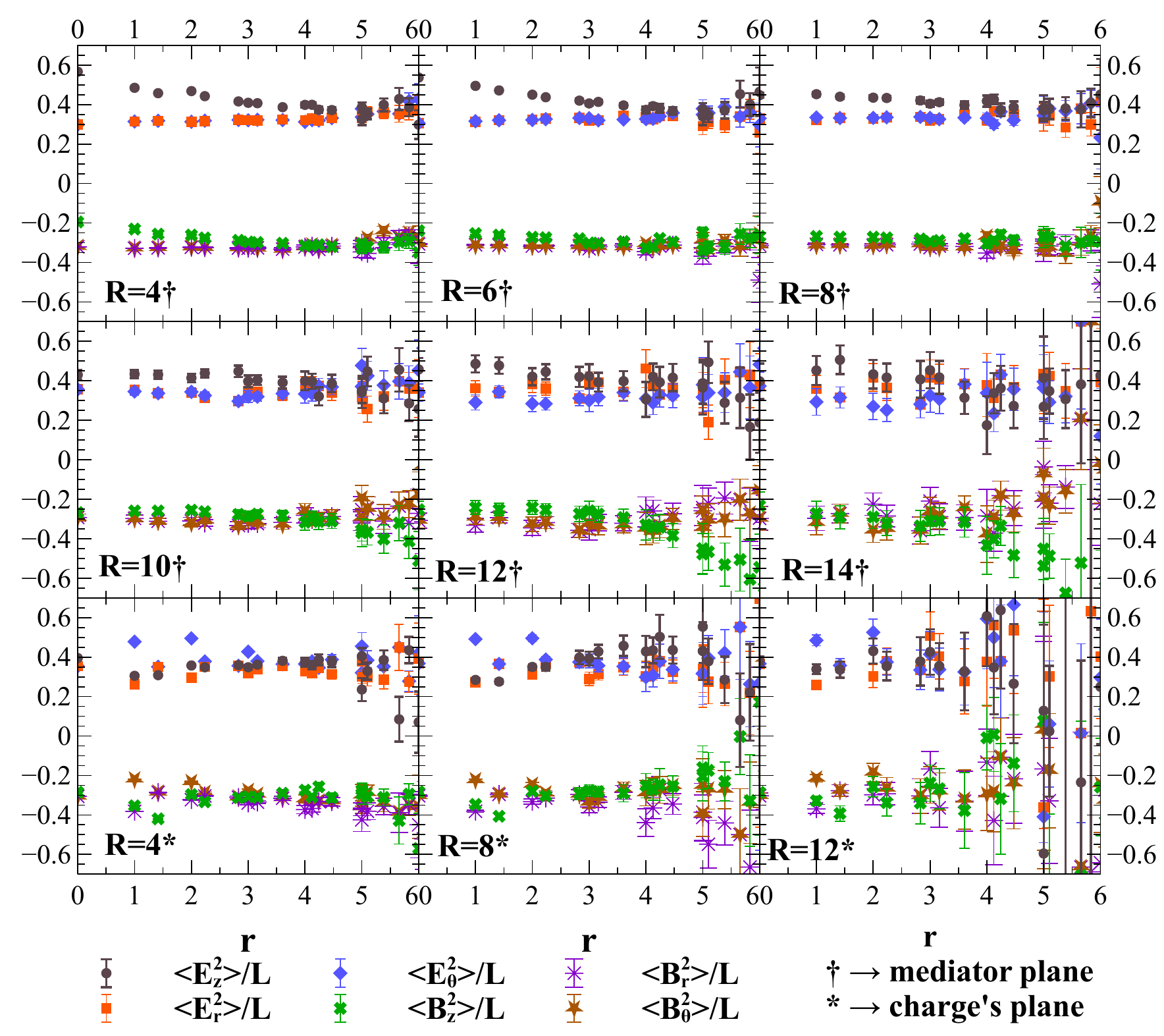}
\end{center}
     \caption{
Ratios of the components of the squared fields over the Lagrangian density, for different inter-charge distances $R$.
We show the ratios both computed for the field profiles computed in the mediator plane of the colour charges and in the planes of the colour charges.
\label{fig:fields_act_ratios}
}
\end{figure*}

\section{Our ansatze and the separation of penetration length, coherence length and quantum widening}

\subsection{In the mediator plane of the two static charges}

In a quantum flux tube, as in the QCD flux tube, at least three parameters, with the dimension of a length, determine the flux tube profile in the mediator plane of the two static charges.   

The quantum width $\omega$ is a function of the flux tube length $R$ and measures the widening of the flux tube due to the zero mode quantum oscillations of the string-like flux tube.

Moreover the flux tube is not an ideal string, and it is due to the squeezing of the fields by the colour confinement. This squeezing is expressed with two parameters. The penetration length $\lambda$ quantifies the exponential screening of the fields penetrating the medium.  But the flux tube cannot just be parametrized by the penetration length, because it should be differentiable at the centre of the flux tube,  with a finite curvature. The coherence length $\xi$ is related to the curvature of the field intensity in the centre of the flux tube.  The penetration length and the coherence length are characteristic of the medium (QCD in our case) where the flux tube resides, and relate to the string tension $\sigma$. They should ideally be measured when the string oscillations are frozen. For instance, the penetration length $\lambda$ and the coherence length $\xi$ are well defined in confinement models such as the Ginzburg-Landau and Amp\`ere  \cite{Cardoso:2010kw} equations or in the Bogoliubov-de Gennes equations \cite{Cardoso:2006mf}.  Notice in these two models for the magnetic confinement in superconductors, the electromagnetic fields are approximated as classical fields, and there is no quantum  widening of the flux tube.

\begin{figure*}[t!]
\begin{center}
\includegraphics[width=1.0\textwidth]{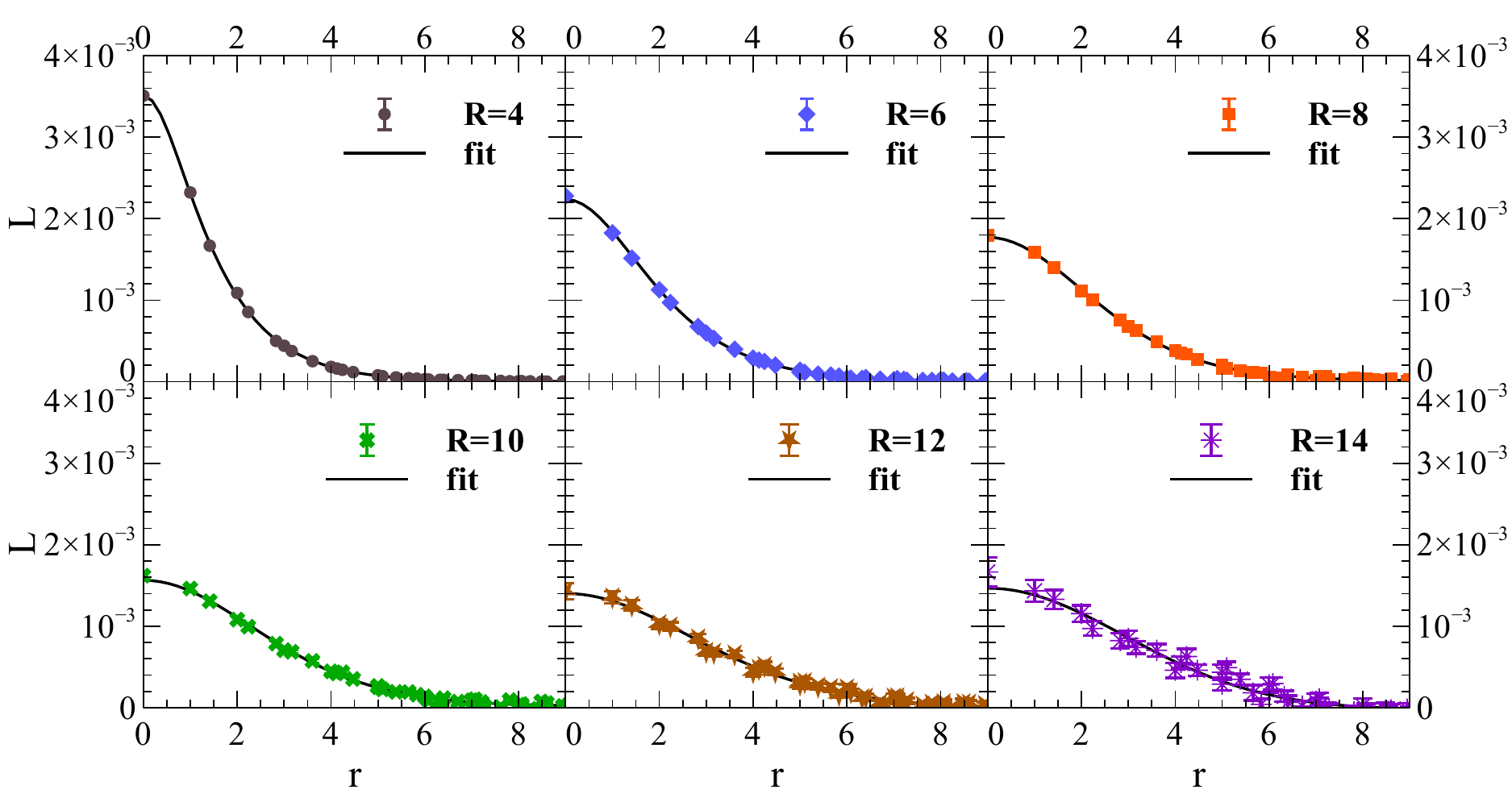}
\caption{
Results for our  fits to the profile of the action density in the mediator of the charges, for $R=4, \ R=6, \ R=8, \ R=10, \ R=12,$ and $ R=14$.
}
    \label{fig:mediator_data_fits}
\end{center}
\end{figure*}

However, here we submit that with present lattice QCD data we can at most fit two lengths in the flux tube profile, because the quantum widening and the classical width are difficult to separate. 
We utilize an ansatz for the flux tube profile to illustrate this difficulty, and we work with the our convention for the cylindrical coordinates is $(r, \theta, z)$. Notice in Fig. \ref{fig:exponent} how the logarithm of the fields we compute is similar to a parabola at small distances and to a line at large distances. Thus  our flux tube profile ansatz has an exponent inspired in the relativistic kinetic  energy, in order to interpolate between a gaussian at small $r$ and an exponential decay at large $r$, 
\bea
\label{ansatz}
{F}^2(r)
& = & 
{F_0}^2 \exp \left(-{2 \over \lambda} \sqrt{r^2 + \nu^2} +  2 {\nu \over \lambda} \right) \ ,
\\ \non
& = &
{F_0}^2 \left[ 1 -  {  r^2 \over \lambda \nu} + o \left( r^4 \over \lambda \nu^3 \right)  \right] \ ,
\\ \non
& = &
{F_0}^2 \exp \left(  {2 \nu \over \lambda} \right)  \exp\left[{-\frac{ 2 r}{ \lambda}} + o\left( \lambda  \over r \right) \right] \ ,
\eea
where $F$ corresponds to any of the components of the squared electric or magnetic fields ${E_r}^2, \, {E_\theta}^2, \, {E_z}^2, \, {B_r}^2, \, {B_\theta}^2, \, {B_z}^2$ or to the lagrangian density $L$.
 Our ansatz is depicted in Fig. \ref{fig:screening-convoluting}, and is expanded for large $r$ and for small $r$ in Eq. (\ref{ansatz}). 
Our ansatz is parametrized with three parameters: the flux tube central intensity $ {F_0}^2$, the flux tube damping measured by the penetration length $\lambda= 1/\mu$ and the flux tube central curvature radius $- 2 {F_0}^2  / ( \lambda \nu )$. In the Ginzburg-Landau case, the curvature is to the coherence length $\xi$. For a simple notation, we utilize as our third parameter the effective distance $\nu$. Notice  the penetration length dominates at large distances, no matter how much curvature we have at the origin.

Let us then consider a typical classical flux tube profile  ${E_{cl}}^2(r)$ as a function of the distance $r$ to the charge axis, similar to our ansatz in Eq. (\ref{ansatz}). Let us convolute the classical flux tube profile with a gaussian distribution, typical of the quantum oscillation,
\be
\phi^2(r)= \exp \left(-{ r^2 \over  \alpha^2} \right)  \ .
\ee
Notice this gaussian already has a width of $w= \sqrt{ \langle r^2\rangle}={ \alpha / \sqrt 2}$.
The result of the convolution is the quantum flux tube profile, 
\be
{F_{qu}}^2(r)= \int_0^\infty \int_0^{2 \pi} \phi^2(r')  {F_{cl}}^2\left(\sqrt{{r'}^2 + r^2 -2 r r' \cos \theta }\right) d \theta \, r' \,  d r' \ .
\label{convolute}
\ee
In Fig.  \ref{fig:screening-convoluting} we also show the numerical result of this integration.

Although we can only compute the integral in Eq. (\ref{convolute}) numerically for all $r$, we are able to compute analytically the profile ${E_{qu}}^2(r)$ both close to the charge axis where the profile quadratic in $r$,
\bea
&&
 {F_{qu}}^2(r)={F_0}^2 \left[
 1-{ \alpha \sqrt{\pi } \over  \lambda}
e^{\left(\frac{\nu}{\alpha}+\frac{\alpha}{ \lambda}\right)^2}
 \text{erfc}\left(\frac{\nu}{\alpha}+\frac{\alpha}{
\lambda}\right)
\right]
\non \\ 
&&
\Biggl\{
1- \Biggl[-
\frac{ \alpha\lambda^2+2 \lambda \nu^2+ 2 \alpha^3}
{ \alpha\lambda^2}
+ \frac{ 2 \nu^2+\lambda\alpha}
{ \lambda \alpha}
\\ \non
&&
{ 1 \over
 1-{ \alpha \sqrt{\pi } \over  \lambda}
e^{\left(\frac{\nu}{\alpha}+\frac{\alpha}{ \lambda}\right)^2}
 \text{erfc}\left(\frac{\nu}{\alpha}+\frac{\alpha}{
\lambda}\right) }
\Biggr] { r^2 \over  \alpha^2} + o \left( r^4 \over \alpha^4 \right)
\Biggr\}
\eea
and at large distances from the charge axis where the penetration length dominates,
\be
 {F_{qu}}^2(r)=
{F_0}^2 \,
\exp\left({\frac{2 \lambda \nu+\alpha^2}{ \lambda^2}}\right) \exp\left[{-\frac{ 2 r}{ \lambda}} + o\left( \lambda  \over r \right) \right] \ .
\ee
These two asymptotic curves to the numerical convolution are shown in Fig.   \ref{fig:screening-convoluting}.
From the result of the convolution, we find that our ansatz is adequate not only for the fit of a classical-like flux tube, but also for the fit of the flux tube with quantum fluctuations, since an interpolation  between the two asymptotic curves yields a very good analytical approximation to the convolution.

\begin{table}[b!]
\caption{\label{tab:fit}
Fits of the profile of the flux tube, for the action density, in the mediator plane for the longitudinal component.  We also consider a constant shift of the density, very small and not shown here}
\begin{ruledtabular}
\begin{tabular}{c|cccc}
$R \ [a]$
&
$10^3 {\cal L}_0 $
&
$\lambda \ [a]$
&
$\nu \  [a]$
&
$ \chi ^2 / dof$
\\
\hline
4 
&
  3.509 $\pm$ 26.72
&
  2.165 $\pm$ 0.033
&
  0.877 $\pm$ 3.335
&
  4.086
\\
6 
&
  2.236 $\pm$ 0.078
&
  2.379 $\pm$ 0.156
&
  2.04 $\pm$ 0.365
&
  2.254
\\
8
&
  1.762 $\pm$ 0.023
&
  2.052 $\pm$ 0.201
&
  4.092 $\pm$ 20.22
&
  1.999
\\
10
&
  1.549 $\pm$ 0.046
&
  2.088 $\pm$ 0.536
&
  5.306 $\pm$ 36.43
&
  1.477
\\
12
&
  1.357 $\pm$ 0.051
&
  0.913 $\pm$ 2.044
&
  17.41 $\pm$ 200.1
&
  1.055
\\
14
&
  1.491 $\pm$ 0.053
&
  0.064 $\pm$ 0.018
&
  268.0 $\pm$ 1392.4
&
 1.331
\end{tabular}
\end{ruledtabular}
\end{table}

Importantly, the penetration length $\lambda$ is unaffected by the convolution, and it is in principle measurable at  the long distance tail of the profile in $r$. 
However it is clear, both from the curvature at the origin and from the radius mean square, that the curvature depends on all three distance parameters $\lambda, \ \nu, \ \alpha$. Thus it is not possible, when error bars are significant, to separate the classical coherence length $\xi$ from the quantum widening $\alpha / \sqrt 2$. 

Moreover, with our ansatz ${F}^2(r)$ defined in Eq. \ref{ansatz}, we obtain the following total width of the flux tube, considering $F^2(r)$ as a distribution function,
\be
{\sqrt{ \langle r^2\rangle}}
 = 
\sqrt{ 
{3 \over 2} \lambda^2 
+ 2 {  \lambda  \nu^2  \over \lambda + 2  \nu } 
} 
\ .
\label{width}
\ee
Thus, our ansatz for the profile in the mediator plane is adequate to study the total width of the flux tube as a function of the inter-charge distance $R$. In the remaining of this paper, we utilize Eq. \ref{ansatz} to fit the profile of the flux tube in the mediator plane, to measure the penetration length $\lambda$ and the total widening $w= {\sqrt{ \langle r^2\rangle}}$.

\subsection{In the planes of the two static charges}

In the planes containing either the quark or the antiquark static charges, only one of the three characteristic distances of the QCD flux tube may be measured. The coherence length is masked by the charges, and the quantum widening only occurs in the flux tube. Thus at most we may measure the screening of the Coulomb field, i e we can only measure the penetration length $\lambda$. 

Nevertheless, for a more detailed study of  the screening, we measure the fields in planes containing one of the two static charges. We compare our lattice data with three different models for the colour fields.  Without confinement, one has a simple Coulomb potential,
\be
 {F_{qu}}^2(r)={F_{0}}^2 { 1 \over r^4} \ ,
\label{source-coulomb}
\ee
when the distance to the charge $r$ is smaller than the inter-charge distance $R$.
If confinement does produce a Yukawa-like screening, the colour fields take the form, 
\be
{F_{qu}}^2(r)={F_{0}}^2  \exp\left( - 2  r \over \lambda \right)  \left( \lambda r + 1 \over r^2 \right)^2 \ .
\label{source-yukawa}
\ee
Finally we may also consider a simple exponential screening similar to the one ocurring in the mediator plane of the flux tube,
\be
{F_{qu}}^2(r)={F_{0}}^2 \exp\left( - 2  r \over \lambda \right)  \ ,
\label{source-screening}
\ee
where $F_{0}$ is just a normalization parameter. 

Then it is important to check whether the penetration length $\lambda$ measured in the plane of the charges is independent of the charge - anticharge distance $R$. For a simple picture of the screening of the colour fields, we must also study if the penetration length $\lambda$ measured in the planes of the charges coincides with the penetration length measured in the mediator plane. 

Thus we measure the colour electric and colour magnetic fields in planes including the charges. Because we consider long flux tubes, we choose to measure the colour fields in the two planes parallel to the mediator plane. These planes are perpendicular to the $z$ axis, and again the variable measuring the distance is $r=\sqrt{y^2 + z^2}$.

\begin{center}
\begin{figure}[t]
\begin{centering}
 \includegraphics[width=0.5\textwidth]{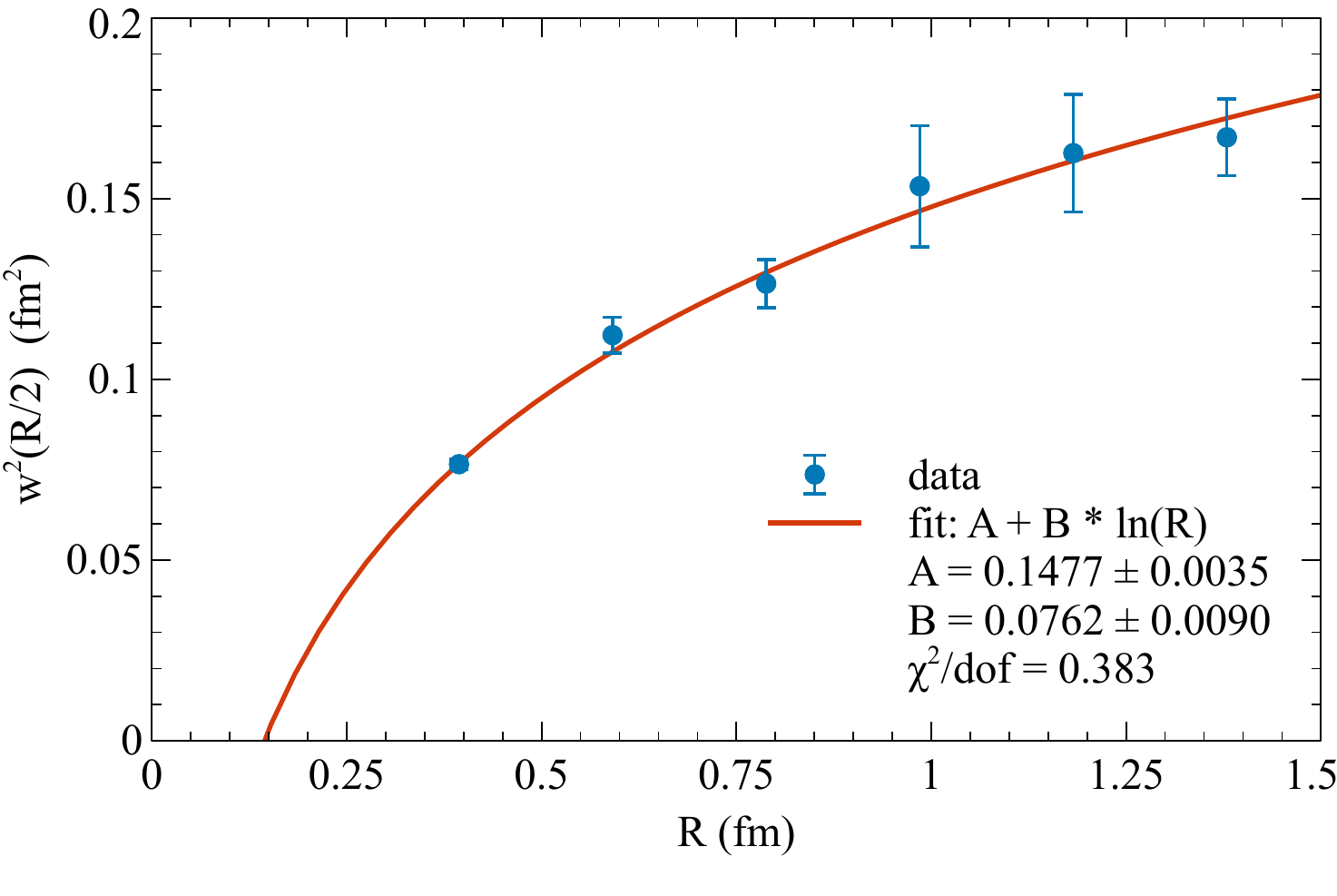}
\par\end{centering}
\caption{
\label{fig:mediator_results}
Square of the width of the flux tube $w^2= \langle r^2 \rangle$ in the mediator plane, computed with our ansatz. The error bars are determined with Jackknife. The solid line corresponds to the fit of the widening of the quantum string.
}
\end{figure}
\par\end{center}

\section{Fits of the flux tube profiles}

\subsection{The squared components of the Electric and Magnetic fields in both planes}

\begin{table}[b!]
\caption{\label{tab:fit_chargesplane}
Parameters of the fits to the profile of the flux tube, for the action density,  in the planes of the charges. We also consider a constant shift of the density, very small and not shown here.}
\begin{ruledtabular}
\begin{tabular}{cccccccccccc}
$R [a]$ &  $10^3 {\cal L}_0 $ & $\lambda[a]$ & $\chi^2/dof$\\ \hline
4 &  5.3917  $\pm$ 17.468  & 2.1088 $\pm$ 0.1212 & 4.8315\\
6 & 4.3832  $\pm$ 20.748  & 2.4803 $\pm$ 0.1376 & 2.1892\\
8  & 4.2056  $\pm$ 11.041  & 2.6118 $\pm$ 0.1788 & 0.9665\\
12& 5.6257  $\pm$ 36.337  & 2.2695 $\pm$ 0.5437 & 2.5743
\end{tabular}
\end{ruledtabular}
\end{table}

\begin{figure*}[t]
\includegraphics[width=1.0\textwidth]{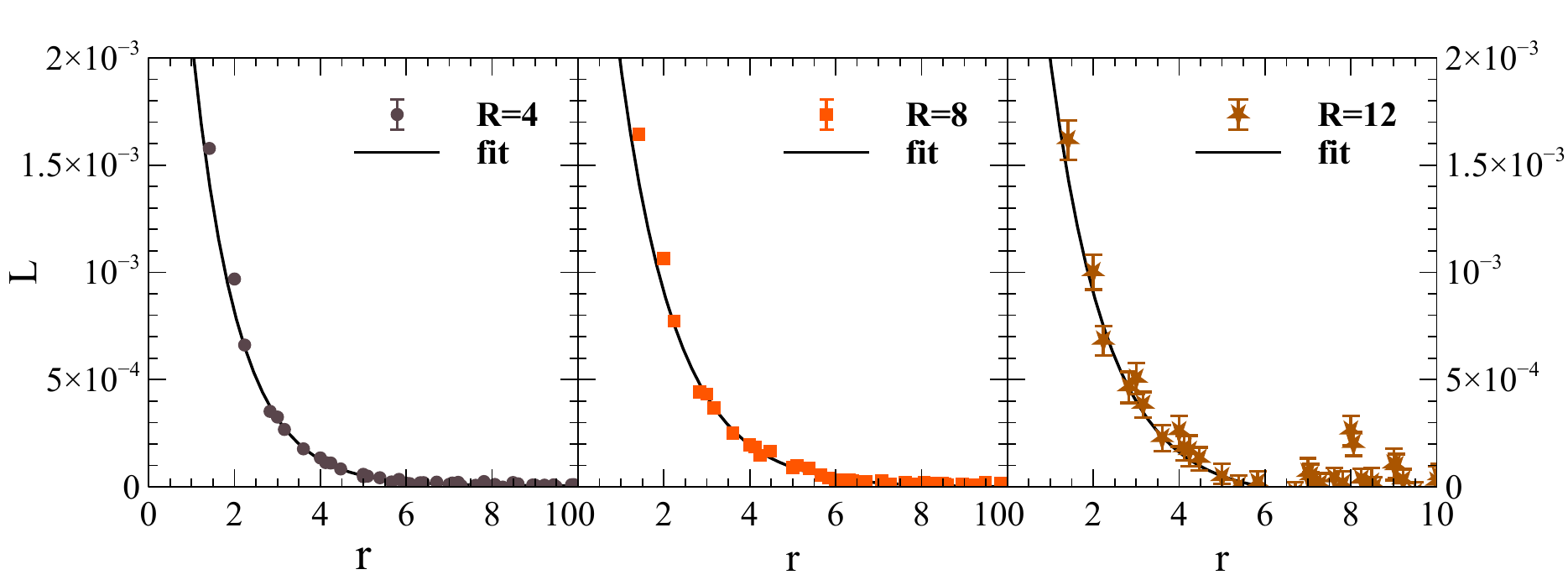}
\caption{
\label{fig:sources_results}
Lattice QCD data and fits with the exponential decay ansatz for the profile of the action density in the planes of the charges, shown  for $R=4$,  $R=8$ and $R=12$.
}
\end{figure*}

Among all densities we measure, the lagrangian or action density is the one with the strongest and clearest signal, therefore this is the density we utilize to parametrize the profiles of the flux tube. Nevertheless all the components squared of the electric and magnetic fields 
${E_z}^2$, ${E_r}^2$, ${E_\theta}^2$, ${B_z}^2$, ${B_r}^2$ and ${B_\theta}^2$, 
are relevant to understand confinement.

In Fig. \ref{fig:fields_act_ratios} we show that, contrary to the dual superconductor models, all
components of the fields are of the same order of magnitude inside the
flux tube. Only close to the charges, the larger component is ${E_z}^2$ in
the mediator plane and ${E_\theta}^2$ in the planes of the charges.

When the distance from the charges is sufficiently large, all the
components ${E_i}^2  \sim 0.4$ and all the components ${B_i}^2 \sim -
0.3$ in lattice spacing units. In any case there is no dominant
component of the colour electric or magnetic fields. This is an
important result that any model of confinement should address.

This also implies that, at sufficiently large distances from the charges, 
the parameter $\lambda$ and the width $w$, computed with any of our field densities,  
are essentially the same .

\subsection{Screening in the mediator plane}

We find the noise increases with $R$ and thus we are able to compute the flux tube profiles only up to $R=14\, a$. We think that our noise suppression techniques are nevertheless sufficient, since the lattice artefacts create larger systematic errors than the statistical noise, see Fig. \ref{fig:artifacts}. As a word of caution we notice the systematic errors may contribute to increase the $\chi^2/$dof .

The fits of the profile of the flux tube in the mediator plane for the action density are shown in Fig.  	\ref{fig:mediator_data_fits} and are listed in Table \ref{tab:fit}.  Notice we only consider in the error bars the statistical error, which increases with $R$, thus decreasing the $\chi^2/ dof$ with $R$. In the smallest distance $R=4$ the systematic errors are larger than the statistical errors, and the  $\chi^2/ dof$ is large. In the largest distances $R=12$ and $R=14$ the statistical errors are already large, and the profile parameters are not well determined. Nevertheless we keep this distance in our study, since the error in the width of the flux tube remains small up to $R=14$.

We remark that, although the other parameters change with $R$, the penetration length $\lambda$ remains the same $\lambda \sim 2.2 \, a$, or  $\lambda \sim 0.22$ fm, within the statistical error bars. This unique scale for the penetration length is promising for the theoretical understanding of confinement.  

The Lagrangian density in the centre of the flux tube and for our largest $R$ is of the order of $1.5 \times 10^{-3}$ in dimensionless units. To arrive at physical dimensions we have to multiply this by 
$2 \beta / a^4 = 2.5 \times10^4 $ GeV  fm$^{-3}$, and we arrive at a Lagrangian density of ${\cal L}_0 \sim 38 $ GeV  fm$^{-3}$. 

\subsection{Widening in the mediator plane}

Since our ansatz fits quite well the flux tube profile, we then utilize Eq. (\ref{width}) to  compute the width of the flux tube. 
Besides, we also compute the error bar or the width with the jacknife, method.
Our results for the width of the flux tube  in the mediator plane are shown in  Fig. \ref{fig:mediator_results}.
As can be seen the tube flux becomes wider as the quark-antiquark
distance is increased. 
We then fit the flux tube width with the leading order one-loop computation in effective string theory
\cite{Gliozzi:2010zt}, corresponding to the linear fit,
\be
w^2 = A + B \log R \ .
\ee
The fit results in, $A = 0.1477 \pm 0.0035$ fm$^2$ and $B= 0.0762 \pm 0.0090$ fm$^2$ with error bars computed with jackknife. Notice the error bars of the fit of the widening, for our larger $R$, are much smaller that the error bars of the parameters $\lambda$ and $\nu$ of our ansatz. Nevertheless we find a rather small $\chi^2/dof = 0.383$. 

The $B$ parameter can be compared with the theoretical leading order  \cite{Gliozzi:2010zt} value for the factor of the logarithmic term,
\be
B= {D-2 \over 2 \pi \sigma}= 0.0640028 fm^2
\ee
obtained using a string tension  of $\sqrt \sigma= 0.44$ GeV \cite{Edwards:1997xf}. 

In what concerns the constant $A$ parameter, since it is positive, it is possibly larger than the corresponding constant of the leading order expansion of the string theory. Possibly this happens since the QCD flux tube is not tachyonic and it's width is always real and positive. Notice a simple exponential profile,  according to Eq. (\ref{width}), already leads for very small distances  to $w^2 = 3 \lambda^2/ 2 \sim 0.07 fm^2$. Indeed this is similar to the width we get at our smaller distance of $R=4a\simeq 0.4$ fm. 

To comply exactly with the quantum widening of an infinitely thin string, the string should be much thinner than longer, and also much thinner than the width of the quantum vibrations. Indeed we have $R >> \lambda$, however $w \sim \lambda$.  That our fitted factor to the logarithm is close to one standard deviation from the theoretical 1-loop result, considering a large part or the width is due to the penetration length, is already a very interesting result.

\subsection{Screening in the planes of the two static charges}

We find that only one of the three ansatze in Eqs.
(\ref{source-coulomb}),
(\ref{source-yukawa})
and
(\ref{source-screening})
fits correctly the action density in the planes of the charges.
Both the Coulomb and Yukawa fields produce very poor fits of our lattice data for the fields.
A poor fit by the Coulomb ansatz was expected since a flux tube is consistent with colour screening. 
However the Yukawa ansatz also leads to a poor fit, and this indicates that the screening occurring in confinement differs from a Yukawa screening.

Importantly, the exponential ansatz fits correctly the tail of the fields in the planes of the charges, see Fig. \ref{fig:sources_results}. Thus we have screening, though it is not a Yukawa screening. Moreover the fit results in a parameter $\lambda \sim 0.22$  to 0.24 fm, as listed in Table \ref{tab:fit_chargesplane} . The $\lambda$ fitted in the planes of the charges is consistent with the $\lambda$ obtained in the mediator plane to the charges.

\section{Conclusions}

We compute the quark-antiquark flux tube in pure gauge SU(3) lattice QCD. We measure the profile of the electric and magnetic field densities both in the mediator plane of the colour charges and in the planes of the charges. We utilize three complementary techniques to enhance the signal to noise ratio, and are able to reduce the statistical noise below the systematic errors of  our lattice setup.

We show the flux tube is due to screening of the electric and magnetic field components, since we measure a penetration length $\lambda \sim 0.22$ to 0.24 fm. The inverse of $\lambda$ may indicate an effective screening mass, possibly for the gluon or dual gluon, of $\mu \sim 0.8 $ to 0.9 GeV. Moreover the same screening parameter is universal in the sense it occurs in all components squared of the electric and magnetic fields 
${E_z}^2$, ${E_r}^2$, ${E_\theta}^2$, ${B_z}^2$, ${B_r}^2$ and ${B_\theta}^2$, both in the mediator plane and in the charge's plane. 

However there are differences to the dual superconductor models. The vector electric and magnetic fields are not gauge invariant, their squared components are the first gauge invariant function of the field components. Moreover, all the squared components have the same order of magnitude, and essentially similar profiles, thus the longitudinal colour electric field is not dominant. 

Importantly, this allows us to use the lagrangian density, since it has the largest signal to noise ratio, to determine the width of the flux tube up to a distance of 14 lattice spacings. We find that the width complies, almost within one standard deviation, with the logarithmic widening obtained at leading order in the Nambu-Gotto effective string theory. 

Our results lead to a better understanding of the nature of the confining SU(3) flux tube. We hope this work will be useful for the theoretical understanding both of the QCD confinement and of string theory.

\acknowledgements
We thank Martin L\"{u}scher,  Uwe-Jens Wiese, and Pedro Sacramento for enlightening discussions on flux tubes.
This work was supported by Portuguese national funds through
FCT - Funda\c{c}\~{a}o para a Ci\^{e}ncia e Tecnologia, projects PEst-OE/FIS/UI0777/2011,
CERN/FP/116383/2010 and CERN/FP/123612/2011. 
We also acknowledge NVIDIA support with an Academic Partnership Program and a CUDA Teaching Center Program.


\bibliographystyle{apsrev4-1}
\bibliography{fluxtube}

\end{document}